\documentclass[preprint, 12pt]{elsarticle}
\usepackage{bm, color, amssymb, lineno, amsmath, booktabs,siunitx,subfigure,mathtools,hyperref}
\usepackage{graphicx}
\usepackage{natbib}
\usepackage{epstopdf,epsfig}

\newcommand{\bi}{\bm}

\biboptions{numbers,sort&compress}

\begin{document}

\begin{frontmatter}
\title{Competing forces in liquid metal electrodes and batteries}

\author[UoR]{Rakan F.\ Ashour}
\author[UoR]{Douglas H.\ Kelley}
\author[hzdr,monterrey]{Alejandro Salas}
\author[hzdr]{Marco Starace}
\author[hzdr]{Norbert Weber}
\author[hzdr]{Tom Weier}

\address[UoR]{University of Rochester, 235 Hopeman Building,
  Rochester, NY 14627, USA}

\address[hzdr]{Helmholtz-Zentrum Dresden -- Rossendorf, %
Bautzner Landstr.\ 400, Dresden, Germany}


\address[monterrey]{Instituto Tecnol\'ogico y de Estudios Superiores de Monterrey, Monterrey, Mexico}
\begin{abstract}
Liquid metal batteries are proposed for low-cost grid scale energy storage. During their operation, solid intermetallic phases often form in the cathode and are known to limit the capacity of the cell. Fluid flow in the liquid electrodes can enhance mass transfer and reduce the formation of localized intermetallics, and fluid flow can be promoted by careful choice of the locations and topology of a battery's electrical connections. In this context we study four phenomena that drive flow: Rayleigh-B\'enard convection, internally heated convection, electro-vortex flow, and swirl flow, in both experiment and simulation. In experiments, we use ultrasound Doppler velocimetry (UDV) to measure the flow in a eutectic PbBi electrode at \SI{160}{\celsius} and subject to all four phenomena. In numerical simulations, we isolate the phenomena and simulate each separately using OpenFOAM. Comparing simulated velocities to experiments via a UDV beam model, we find that all four phenomena can enhance mass transfer in LMBs. We explain the flow direction, describe how the phenomena interact, and propose dimensionless numbers for estimating their mutual relevance. A brief discussion of electrical connections summarizes the engineering implications of our work. 
\end{abstract}

\begin{keyword}
liquid metal battery \sep electro-vortex flow \sep internally heated
convection \sep swirl flow \sep Rayleigh-B\'enard convection
\end{keyword}

\end{frontmatter}
\clearpage

\section{Introduction}
Electrical grids work by balancing the level of demand and
supply. Today, the fluctuations are typically handled by gas-fired 
turbines, which are turned on during the peak demand of electricity.
The electrical grid must be designed for that peak power since
it lacks an efficient storage mechanism to dampen the
load. The increasing integration of renewable energy sources will
likely cause even higher fluctuations that cannot be handled by the grid in
its current state \citep{Backhaus2013,IEA2013}. Large scale stationary
energy storage would likely become mandatory to compensate the fluctuations.
Due to its low price, the liquid metal battery (LMB) might be an ideal
storage technology \citep{Kim2013b,Wang2014}. 

\begin{figure}[bth]
\centering
\subfigure[]{\includegraphics[height=5cm]{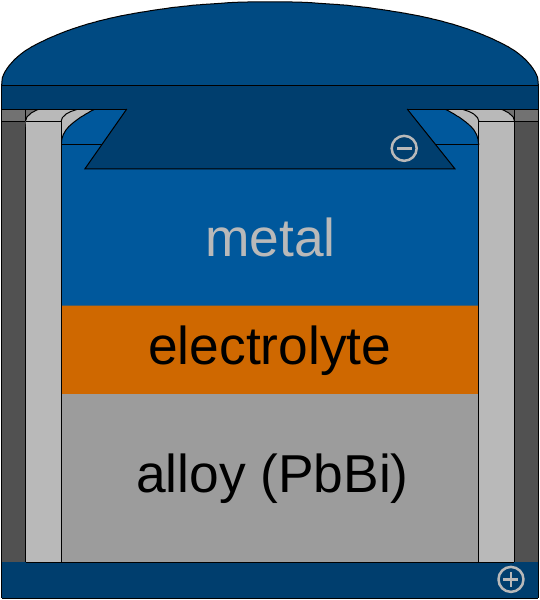}}\hfill
\subfigure[]{\includegraphics[height=5cm]{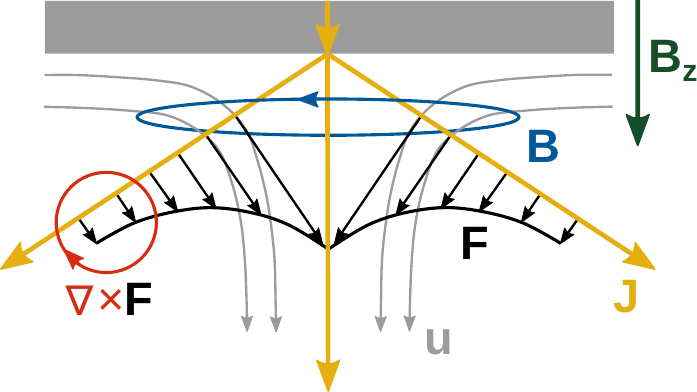}}
\caption{Sketch of a typical liquid metal battery, which is cylindrical and shown in cross-section (a), and electro-vortex flow at a point current source (b). Diverging current produces a force with nonzero curl that necessarily drives flow, and current always diverges from the negative current collector of a liquid metal battery.} 
\label{f1}
\end{figure}

An LMB is composed of two liquid metals with different
electronegativity separated by a layer of molten salt electrolyte
(Fig.~\ref{f1}a). Typical chemistries include Ca$||$Bi
\cite{Kim2013a,Ouchi2016}, Ca$||$Sb \cite{Poizeau2012,Ouchi2014},
K$||$Hg \cite{Agruss1967}, Li$||$Bi
\cite{Foster1964,Ning2015},
Li$||$Cd \cite{Cairns1967}, Li$||$Pb \cite{Cairns1967,Wang2014},
Li$||$Sb \cite{Wang2014}, Li$||$Se
\cite{Cairns1969b,Cairns1969c}, Li$||$Sn
\cite{Foster1966b,Hesson1967}, Li$||$Te
\cite{Shimotake1969},
Li$||$Zn \cite{Cairns1967}, Mg$||$Sb \cite{Bradwell2012},
Na$||$Bi
\cite{Shimotake1967,Foster1967a},
Na$||$Hg \cite{Heredy1967,Spatocco2014}, Na$||$Pb
\cite{Cairns1967}, Na$||$Sb
\cite{Kim2013b}, Na$||$Sn
\cite{Weaver1962,Agruss1963,Agruss1967}, Na$||$Zn \cite{Xu2016,Xu2017} 
and Zn$||$Bi,Sn,Pb \cite{Holubowitch2016,Holubowitch2016b} cells. The
three liquid layers are self-segregated based on density, which makes
the manufacturing process simpler and less expensive compared to other
batteries. We will consider LMBs of cylindrical shape, though rectangular shapes are also possible. When the battery is connected to an external load, alkali
or alkaline metal  ions (A$^{\text{+}}$) are produced via
oxidation. These metal ions have to travel through the molten
electrolyte before they are reduced at the surface of the positive
electrode (B) where they form an alloy A$_{\text{in B}}$. The open
circuit potential of the battery depends on the Gibbs free energy of
the alloying reaction \citep{Kim2013b}. The all-liquid design allows
for fast charge transfer kinetics \cite{Swinkels1971} and current
densities up to 13 A cm$^{-2}$
\cite{Cairns1967,Cairns1969b,Cairns1969c}. At low currents, the 
ohmic losses in the electrolyte layer are the most important
overpotential. However, at higher current densities concentration
polarization becomes the main challenge
\cite{Heredy1967,Agruss1967,Foster1967b,Agruss1963}; due to the slow
diffusion of reaction products \cite{Kim2013b,Bradwell2012},
intermetallic phases are formed at the cathode-electrolyte interface
\cite{Cairns1967,Vogel1967}. Often they float on top of the cathode
metal because of their low density \cite{Ouchi2014}, and sometimes dendrites may even short-circuit the 
cell \cite{Kim2013a}. The solid intermetallic layer will not only hamper mass
transfer, but will also act as a solid wall slowing down the flow in
the cell. A local accumulation of intermetallic phases will lead to a
modified conductivity distribution, causing small scale electro-vortex
flow. Hence, a gentle mixing of the bottom electrode
may crucially increase cell performance.

Fluid flow in individual cells does not only affect the cell's
efficiency. Perhaps more importantly, the development of fast flows
imposes a risk of interface deformations that can lead to an
electrical short and sudden discharge of the cell. Prior work in that context
focused on the Tayler instability (TI)
~\citep{Stefani2011,Seilmayer2012,Weber2013,Weber2014,Herreman2015,Starace2014,Weber2015b,Weier2017}, 
which will appear first in the top electrode. The TI may be suppressed
by different means \cite{Stefani2011,Weber2014} and is likely to short-circuit
only very large cells (with a diameter on the order of meters)
\cite{Herreman2015}. In contrast, magnetohydrodynamic (MHD)
instabilities of the metal-electrolyte interfaces may appear in small
cells, too. Well known from aluminum reduction cells, this
``sloshing'' or ``metal pad roll'' instability can be
explained by the interaction of horizontal currents with a magnetic
background field
\cite{Sele1977,Zikanov2015,Weber2017,Weber2017a}. Small density
differences between metal and electrolyte and strong vertical magnetic
fields promote this instability \cite{Bojarevics2017,Weber2017a}. In
LMBs, the
two liquid interfaces can influence each other and may lead to very
complex coupling \cite{Horstmann2017,Zikanov2017} which is still under
investigation. Further, thermal convection (TC) is expected
to appear especially in the electrolyte and upper metal
layer. Although TC will be present in very small cells, its (moderate)
flow will not endanger the safe operation of small and
medium sized LMBs \cite{Shen2015}. Marangoni convection driven by a surface tension gradient is not 
expected to influence the flow velocity considerably, but is expected to change
the size of the Rayleigh-B\'enard cells
\cite{Koellner2017}. Finally, TC was also found to enhance mass
transfer in the positive electrode \cite{Kelley2014,Beltran2016},
e.g. by heating the cell from below. For a discussion of the thermal balance of 
an LMB, see \cite{Wang2015}.

Electro-vortex flow (EVF)
can also drive flow in LMBs. Its
usefulness for mixing LMBs and its implications
for the safe operation of LMBs were already discussed in
\cite{Bradwell2011a,Bradwell2015} and \cite{Weber2014b}. The origin of EVF
is best explained by the illustrative example of Shercliff
\cite{Shercliff1970} (Fig. \ref{f1}b). An electric current density $J$ passes
from a point source into an infinite fluid layer. Within the fluid, the
current diverges radially. The resulting Lorentz force $F$ (which is
the cross product of the current density with its own magnetic field
$B$) is non-conservative, i.e. its curl does not vanish. It can
therefore not be balanced by a pressure gradient and drives a flow
$\bm{u}$ away from the wall. This phenomenon is commonly denoted as
``electrically driven vortical flow'' or shorter ``electro-vortex
flow''. The term ``vortex'' refers to the typical flow structure
observed in confined volumes. EVF is caused by \emph{internal}
currents producing a \emph{rotational} Lorentz force and driving a jet
away from the wall \cite{Shercliff1970}. It cannot exist in 2D
\cite{Shercliff1970}. Since battery electrodes are typically wider than the wires that supply their currents, and current collectors must be made small enough to prevent shorts with vessel walls, battery currents almost always diverge, at least in some regions of the electrode. The extent to which they diverge and drive electro-vortex flow depends on connection topology. 

Already one year before Shercliff, Lundquist
studied the same problem. He discovered that the velocity of EVF
scales with the so-called EVF parameter $S=\mu_0
I^2/(4\pi^2\rho\nu^2)$ \cite{Lundquist1969,Shcherbinin1975} with
$\mu_0$, $I$, $\rho$ and 
$\nu$ denoting the vacuum permeability, the current, the density, and
the kinematic viscosity, respectively (for different definitions, 
see \cite{Sozou1971,Vlasyuk1987b}; \cite{Vlasyuk1987b} might have a
typo). At typical currents for
industrial applications the velocity scales as $u\sim\sqrt{S}$,
i.e. linearly with the current. Only at low currents, a quadratic
scaling is observed
\cite{Shcherbinin1975,Vlasyuk1987b,Bojarevics1989}. It should be 
further noted that the jet is steady only at very low velocities;
typically, it oscillates (see e.g.,
\cite{Davidson2001,Shatrov2012}). At very high velocities, the flow 
can generally saturate by the induced current or magnetic field 
(as, e.g., the Tayler instability \cite{Weber2015b,Weier2017}). However,
at typical velocities, the main dissipation happens in the boundary 
layers \cite{Davidson1999,Davidson2001} (for an experimental study 
see \cite{Volokhonskii1991}).
The thoroughly studied problem of an
infinite fluid layer
\cite{Zhigulev1960,Zhigulev1960a,Lundquist1969,Sozou1971,Sozou1972,Sozou1972b,Narain1973,Jansen1986,Petrunin1993}
(overview \cite{Shcherbinin1975,Moffat1978}) is therefore of limited
practical relevance because it involves no boundary layers.

Davidson provides a concise, clear, and understandable introduction to electro-vortex flow
\cite{Davidson2001}. For an even shorter introduction, see Chudnovskii
\cite{Chudnovskii1989} and for a broad overview on EVF and its
application Bojarevics et al. \cite{Bojarevics1989}. A number of experiments on
electro-vortex were conducted by Woods et al. \cite{Woods1971}, Butsenieks et al.
\cite{Butsenieks1976,Millere1980,Vlasyuk1987b,Chudnovskii1989b},
Bojarevics et al. \cite{Boyarevich1977}, Zhilin et al.
\cite{Zhilin1986,Volokhonskii1991,Weber2017} and R\"abiger et al.
\cite{Raebiger2014,Zhang2016,Franke2016}; for an excellent
overview, see \cite{Bojarevics1989}.

At low currents the jet illustrated in Fig. \ref{f1}b usually forms a poloidal flow that descends along the axis of the cylindrical electrode and rises along the walls, since it is confined by boundaries. However, at high currents an azimuthal flow or ``spontaneous swirl'' \cite{Davidson2001} is often observed experimentally \cite{Woods1971, Bojarevics1983}. Its origin can be explain similar to the Sele mechanism in aluminum reduction cells \cite{Sele1977,Weber2017a}. A vertical magnetic background field and the radial cell current produce an azimuthal force driving an azimuthal flow. The magnetic background field may be the Earth's magnetic field \cite{Bojarevics1983}; similarly a ``slight deviation from symmetry'' in the experiment will lead to swirl \cite{Boyarevich1977, Davidson1999, Shatrov2012}. Making electrical connections among a tightly-packed array of batteries is almost certain to cause such deviations from symmetry. The spin-up of swirl takes a significant time (around a minute in large vacuum arc remelting vessels) \cite{Davidson2000}; therefore, swirling may be minimized by periodically reversing the direction of the vertical background field \cite{Davidson2000}.

Curiously, the swirl is known to almost totally supersede the poloidal flow. Davidson explains that
``the radial stratification of angular momentum suppresses radial
motion in the same way as density stratification suppresses axial
motion'' \cite{Davidson2000}. He demonstrates that the azimuthal flow
suppresses the poloidal one (``poloidal suppression'') and argues that
Ekman pumping \cite{Schlichting2006} is the key dissipative mechanism
\cite{Davidson1999}. He further predicts that an azimuthal force of
only 1\% of the poloidal one will lead to a strong swirl flow. Alternative explanations have also been proposed, however~\cite{Millere1980,Bojarevics1983,Shtern1995}.

In this article we study the flow in the bottom electrode only. Specifically we investigate the interaction of four different flow phenomena: electro-vortex flow, swirl flow, internally heated convection (IHC) and classical Rayleigh-B\'enard convection. All four depend strongly on the location and topology of electrical connections to the battery. Those connections are likely to be tightly packed into a large array of batteries, since the technology is intended to store enough energy for neighborhoods or cities. Where the connections enter and leave each battery, how much they concentrate the electrical current, and what paths they take outside the batteries can all affect flow in the electrode. To study the effects of connection on the four flow phenomena, we use a combination of experimental and numerical models of PbBi electrodes operating at mean current densities of up to 0.65\,A~cm$^{-2}$. 

The paper proceeds as follows: in the next section we describe the experimental apparatus and numerical model, followed by an order of magnitude estimation of the different flows in section \ref{sec:equations}. Finally we present and discuss the results in section \ref{sec:results}.

\section{Methods}
\subsection{Experimental}
\begin{figure}[t!]
\centering
  \includegraphics[width=\textwidth]{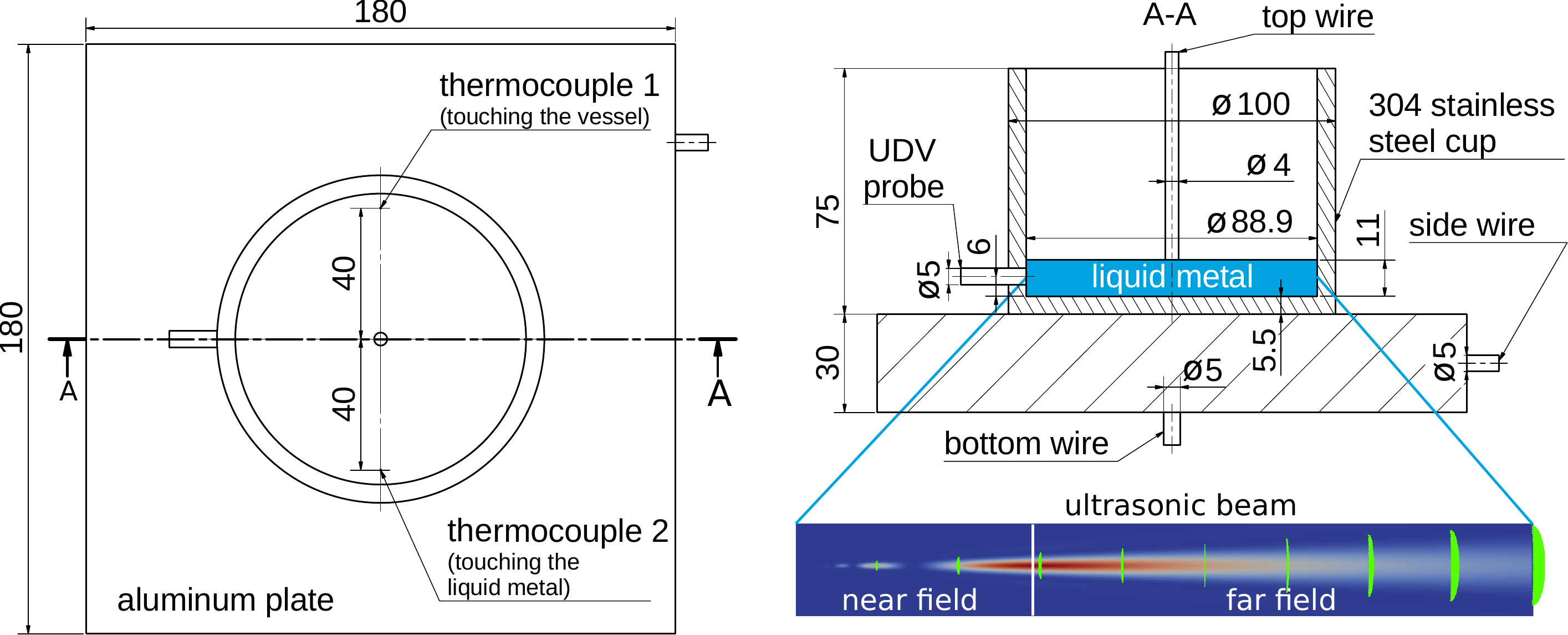}
\caption{Dimensions of the experiment and simulation model (in mm) and
  illustration of the ultrasound beam showing the near and far field
  as well as the beam broadening.}
\label{fig:2}
\end{figure}

A schematic of the experimental apparatus is shown in
Fig.~\ref{fig:2}. The vessel is a cylinder of 304 stainless steel
with inner radius $R=44.5$~mm that also serves as the positive current
collector. The liquid metal used in the experiments is a eutectic
alloy of 44.5\% lead and 55.5\% bismuth. The vessel rests on an 
aluminum plate which is much more conductive than stainless steel,
thereby allowing heat and current density to homogenize before
entering the vessel. The upper current collector is a 4~mm-diameter 
nickel-plated copper wire with mica and fiberglass insulation. It
is mounted on a PTFE sheet using shaft
collars and is connected to a current-controlled DC power source. Two
K-type thermocouples are placed at different depths to measure the
temperature difference between the top and bottom of the liquid metal
electrode. A third thermocouple is connected to a
proportional-integral-derivative controller, which allows for the
temperature to be maintained at \SI{160}{\celsius} (at the bottom of the
cell). The temperature is chosen to match the operating conditions of
a Na$||$PbBi battery~\citep{Ashour2017}. The apparatus
minimizes heat exchange by using ceramic insulation, which allows for
the operating temperature to be maintained for the duration of the
experiment~\citep[see][]{Kelley2014,Perez2015}.

Since liquid metal is opaque, it is not possible to measure flow using
optical methods like particle tracking or particle image
velocimetry. Therefore, to study mixing and transport in liquid metal
electrodes we use an ultrasound Doppler velocimeter (Signal-Processing
DOP 3010). The velocimeter drives an ultrasound probe that makes a
beep and listens for echoes. The distance of an echoing body is
determined from the speed of sound and the time delay between emitted
and received signals. The speed of that body toward or away from the
probe is determined from the Doppler shift of the echo. Negative
values in the velocity profile indicate a flow moving toward the
probe, whereas positive values indicate a flow away from the
probe. Small oxide particles or other impurities common in liquid
metal provide sufficient echoes. The probe operates with a frequency
of 8~MHz and has a piezoelectric emitter of diameter 5~mm, which gives a radial
resolution of 0.5~mm and lateral 
resolution of a few mm. In the apparatus shown in Fig.~\ref{fig:2},
the ultrasound probe is mounted on the side of the cylinder allowing
for the measurement of the radial velocity component of the flow along
a diameter, 6~mm above the vessel floor. The cylinder is filled with
700~g of eutectic PbBi, which gives a realistic electrode thickness of
about 11~mm. Though UDV measurements in similar shallow layers were
already successfully conducted in the past \cite{Kolesnichenko2005},
our experiment is fairly challenging. As the UDV beam broadens on its
way (Fig. \ref{fig:2}), it samples flow in a finite volume in which the velocity is not necessarily uniform; regions of flow towards and away from the probe may
cancel each other out. In order to correctly compare the simulated
velocities, we use a simple UDV beam model (\ref{s:beamModel}).

To study the effect of electric current topology on fluid
flow, two different current paths are investigated. In one arrangement,
current is supplied to the side of the aluminum plate. This model is
very likely to be seen in battery packs (see e.g., Fig. 20 in
\cite{Bradwell2011a,Bradwell2015}). In the other arrangement, current is
supplied to the bottom of the aluminum plate through a 100 cm long
aluminum rod. This arrangement represents vertically stacked cells and minimizes symmetry-breaking in the supply currents, thereby reducing swirl flow.

\subsection{Mathematical and numerical model}
In our simulations, thermal convection and electro-vortex flow are modeled separately. The
two models are implemented in the open source library OpenFOAM
\cite{Weller1998}. If not otherwise
stated, the following material properties are assumed for eutectic
PbBi at \SI{160}{\celsius}: a kinematic viscosity of
$\nu=2.7\cdot10^{-7}$\,m$^2$s$^{-1}$, a volumetric thermal expansion coefficient of
$\beta = 1.3\cdot 10^{-4}$\,K$^{-1}$, an electric conductivity of
$\sigma=9\cdot 10^5$\,S~m$^{-1}$, a density of $\rho = 10505$\,kg~m$^{-3}$, an
isobaric heat capacity of $c_p = 148$\,J~kg$^{-1}$K$^{-1}$, a heat conductivity of
$\lambda = 10$\,W~m$^{-1}$~K$^{-1}$, a thermal diffusivity $\alpha=6\cdot
10^{-6}$\,m$^2$~s$^{-1}$, a Prandtl number of $Pr=0.04$ and a sound velocity
of $c = 1765$\,m~s$^{-1}$ \cite{Sobolev2007,Sobolev2010,NEA2015}. The
electric conductivity of the vessel is assumed to be $\sigma=1.37\cdot
10^6$\,S~m$^{-1}$ and that of the aluminum plate $\sigma = 3.5\cdot 10^7$\,S~m$^{-1}$. 

\subsubsection{Thermal convection}
Thermal effects are modeled in the fluid only. All solid conductors
are replaced by simplified boundary conditions. The flow
is simulated using the Oberbeck-Boussinesq approximation
\cite{Oberbeck1879}. Its validity is estimated according to Gray and
Giorgini \cite{Gray1976}: a temperature difference of 27\,K would lead
to an error of 10\,\% of the dynamic viscosity, when neglecting its
temperature dependence. Respectively, the temperature difference may
reach 66\,K for thermal conductivity and 833\,K for $c_p$ and density
in order to stay below the same error. As we expect temperature
differences in the order of 10\,K, the Oberbeck approximation is
well-suited for our problem. The Navier-Stokes and energy
equations
\begin{eqnarray}
  \frac{\partial \bi u}{\partial t} + \nabla\cdot(\bi u\bi u) &=&
  -\nabla p_d + \nu\Delta\bi u - \bi g\cdot\bi r\nabla\rho_k\\
  \nabla\cdot\bi u &=& 0\\
\frac{\partial T}{\partial t} + \nabla\cdot(\bi u T) &=&
\frac{\lambda}{\rho_0 c_p}\Delta T + \frac{\bi J^2}{\sigma\rho_0 c_p}
\end{eqnarray}
are solved, with $\bi u$, $t$, $\bi g$, $\bi r$, $T$, and $\bi
J$ denoting velocity, time, 
gravitational acceleration, position, temperature, and current
density, respectively. The density
$\rho=\rho_0\rho_k = \rho_0(1-\beta(T-T_{\text{ref}}))$ is calculated using
the mean density $\rho_0$ at reference temperature $T_{\text{ref}}$ and the
coefficient of thermal expansion $\beta$. The current density is
computed with the electro-vortex solver and provided as initial
condition. The modified pressure is defined as
$p_d = (p-\rho\bi g\cdot\bi r)/\rho_0$, where $p$ is the pressure. The set of equations is
solved using the PISO algorithm on a collocated grid (with Rhie-Chow
interpolation \cite{Zhang2004}) employing a modified version of the
OpenFOAM standard solver \emph{buoyantBoussinesqPimpleFoam} and using
at least 300 cells on the diameter. 

\subsubsection{Electro-vortex flow}
Simulating electro-vortex flow requires the coupling of solid and
liquid conductors to obtain the correct current density. This coupling is achieved by a parent-child mesh
technique. While electric properties (electric potential, current
density) are solved on a global mesh, the flow is computed in the
fluid region only, so quantities must be mapped between meshes. 
The magnetic field is determined by an
integro-differential approach, i.e.\ its boundary conditions are
computed using Biot-Savart's integral \cite{Weber2013} and the
induction equation is solved in the fluid. The model is described in
detail in \cite{Weber2017b}.

The electric potential $\varPhi$, current density $\bi J$ and magnetic field
$\bi B$ are split into constant (subscript 0) and induced parts (lower
case) as
\begin{eqnarray}
\varPhi &=& \varPhi_0 + \varphi\\
\bi J &=& \bi J_0 + \bi j\\
\bi B &=& \bi B_0 + \bi b.
\end{eqnarray}
The induced magnetic field $\bi b$ is neglected. The following set of
equations is solved on the global mesh:
\begin{eqnarray}
  \nabla\cdot\sigma\nabla\varPhi_0 &=& 0\\
  \nabla\cdot\sigma\nabla\varphi &=& \nabla\cdot \sigma(\bi u\times\bi B)\\
  \bi J_0 &=& -\sigma\nabla\varPhi_0\\
  \bi j &=& \sigma(-\nabla\varphi + \bi u\times\bi B).
\end{eqnarray}
In the fluid region we solve
\begin{eqnarray}
&&\frac{\partial \bi u}{\partial t} + \left( \bi u \cdot \nabla \right)
\bi u = - \nabla p + \nu \Delta \bi u + \frac{\bi J\times\bi
  B}{\rho}\\
&&\bi B_0(\bi r) = \frac{\mu_0}{4\pi}\int \frac{\bi J_0(\bi
  r')\times(\bi r
  - \bi r')}{|\bi r - \bi r'|^3}dV'\\
&&0=\Delta\bi B_0
\end{eqnarray}
with $\mu_0$ and $V$ denoting the
vacuum permeability and cell volume, respectively. Computing Biot-Savart's
integral, the current $\bi J$ of the full mesh is used for determining
$\bi B$ at the fluid boundaries only. The grid resolution is
approximately 100 cells on the diameter with strongly refined 
boundary layers.

\section{Theory of transport in liquid metal
  electrodes}\label{sec:equations}
In this section, we will derive two simple dimensionless numbers for
estimating which flow dominates: thermal convection or 
electro-vortex flow. We will start with classical Rayleigh-B\'enard
convection.

A typical liquid metal battery setup requires heating a
cylindrical container from the bottom, which generates a temperature
gradient along the depth of the electrode. As the liquid closer to the
bottom becomes less dense, it rises to the surface allowing for the
cooler fluid to sink down. A steady flow is established when the
balance of kinetic energy and viscous dissipation is reached. In
systems where the viscosity is high enough, or length scales are small
enough, fluid motion can be completely suppressed. The onset of
thermal convection is defined by the Rayleigh number (i.e. the ratio of
buoyant to viscous forces) as
\begin{eqnarray}\label{Rayleigh_num}
Ra = \frac{g\beta h^3 (T-T_0)}{\alpha\nu},
\end{eqnarray}
with $g$, $h$, and $T-T_0$ denoting the
gravitational acceleration, the
height of the layer, and the temperature difference, respectively, all at working temperature $T_0 = \SI{160}{\celsius}$.
In our specific experiment with a free upper surface,
convection is expected to set in at a critical Rayleigh number of
$Ra_\text{cr}=1100$ \cite{Kundu2012}, which corresponds to a temperature
difference of $\Delta T \approx 1$\,K. We expect therefore strong
Rayleigh-B\'enard convection in the experiment, which we maintain at a
temperature difference of $\Delta T \approx 8$\,K. The typical velocity scale of
Rayleigh-B\'enard convection can be estimated by the free fall velocity
as \cite{Shi2012,Shen2015}
\begin{eqnarray}\label{eq:ScalingBuoyancy}
U_b \sim \sqrt{g \beta \Delta T h}.
\end{eqnarray}

We will now proceed with the case of internally heated convection
(IHC), which is
caused by Joule heating of the electric current due to ohmic losses. Similarly, the
free fall velocity may be used as velocity scale. However, some
characteristic temperature difference needs to be defined
first. Following Goluskin \cite{Goluskin2015,Shen2015} we can
use the heating rate $\dot Q' = J^2 / (\sigma\rho c_p)$ \cite{Kazak2013}
to obtain a characteristic temperature gradient
\begin{eqnarray}\label{eq:dT}
  \Delta T = \frac{h^2 J^2}{\alpha \sigma \rho c_p}.
\end{eqnarray}

Finally, we consider electro-vortex flow. When an electrical current
density $\boldsymbol{J}$ is injected into the fluid, it induces a
magnetic field $\boldsymbol{B}$, and the two are related via Ampere's
law as
\begin{eqnarray}\label{eq:Ampere}
    \oint_C {\boldsymbol{B}\, d\ell = \mu _0\int_S \boldsymbol{J}\,dS},
\end{eqnarray}
with $S$, $C$ and $dl$ denoting a surface, the curve around the
surface and an infinitesimal element of the curve.
In geometries relevant for liquid metal batteries, electrical current
flows between two current collectors, both aligned on the axis of
symmetry. Typically, the upper current collector has a smaller diameter
than the lower one. Given that geometry, we first consider purely
axial current, which would drive a circulation that rises near the
side walls and descends near the center. To see why, apply
(\ref{eq:Ampere}) to two paths of integration, both circles of
the same radius and both aligned with the central axis, but at
different heights. Since current is more concentrated near the
smaller, upper current collector, the magnetic field there is
larger. The Lorentz force points toward the central axis everywhere;
however, it has larger magnitude near the upper current collector,
causing an inward flow there. Conservation of mass then requires
outward flow near the lower current collector, sinking near the
center, and rising near the side walls. Arguments made with this 
poloidal circulation in mind lead to an expected scaling of
electro-vortex flow as~\citep{Davidson2001}
\begin{eqnarray} \label{eq:ScalingLorentz}
U_L \sim \frac{(\mu _0/\rho)^{1/2} I}{2 \pi R},
\end{eqnarray}
with $\mu_0$, $I$ and $R$ denoting the vacuum permeability, the cell
current and radius, respectively. For other (but similar) possible
scales, see
\cite{Vlasyuk1987b,Chudnovskii1989b,Bojarevics1989,Kazak2012}.

In a flow in which both buoyant and electromagnetic forces are
present, we expect flow speed to be predicted by
(\ref{eq:ScalingBuoyancy}) when buoyancy dominates and
by (\ref{eq:ScalingLorentz}) when electromagnetic
forces dominate. We can moreover estimate which of the two effects 
dominates. The ratio of the characteristic velocities of electro-vortex flow to Rayleigh-B\'enard
convection is
\begin{eqnarray}\label{eq:A}
A \sim \frac{(\mu _0/\rho)^{1/2} I}{{2 \pi R}~\sqrt{ g \beta \Delta T
    h}}
\end{eqnarray}
and the ratio of the characteristic velocities of electro-vortex flow to internally heated convection
\begin{eqnarray}\label{eq:B}
B\sim \frac{R}{2h}\sqrt{\frac{\mu_o \sigma \lambda}{\rho g \beta h}}.
\end{eqnarray}
For alternative expressions, see \cite{Atthey1980,Nikrityuk2007,Kazak2013}.
We can think of equation (\ref{eq:A}) as the ratio of momentum gained
from electromagnetic forces to that gained from buoyant forces in the core of the
electrode, whereas equation (\ref{eq:B})  can be thought of as the
ratio of momentum gained by electromagnetic forces to that dissipated
by Joule heat. Hence, when $A \ll 1$ or $B \ll 1$, we expect buoyancy to
dominate the flow. When $A \gg 1$ and $B \gg 1$, we expect electro-vortex flow
to dominate.

The dimensionless ratios $A$ and $B$ are obtained from a simple
formalistic approach. Compared to the sophisticated considerations of
Davidson et al. for a similar problem \cite{Davidson2000}, our approach is
 quite simplistic~--- probably too simple. Most notably, our
formulas do not cover the \emph{distribution} of the current density
nor the diameter of the current collectors. These will~--- of course~---
have a crucial influence on the flow magnitude, though they will not change $A$ and $B$. Nevertheless, we
believe it is admissible to use $A$ and $B$ as a first estimate for
the relative importance of thermal convection and electro-vortex
flow. Table \ref{table:1} shows both dimensionless quantities for our
geometry for eutectic PbBi as well as liquid sodium. Considering the
low values of $A$, we expect Rayleigh-B\'enard convection to dominate in
the experiment. This is indeed the case, as we will show later. On the
other hand, the dimensionless quantity $B$ is clearly larger than
one. Thus, we expect electro-vortex flow to dominate over
internally heated convection. In experiments, however, the unavoidable presence of classical 
Rayleigh-B\'enard convection does not allow comparing EVF and IHC 
directly.

Finally, a third dimensionless ratio can be used to estimate the
transition from poloidal electro-vortex to an azimuthal swirl flow. As
its derivation is effectively explained by Davidson et al. 
\cite{Davidson1999,Davidson2001}, we give here only the result: if an
azimuthal force of a magnitude of 1\,\% of the poloidal one is
present, we expect swirl flow to dominate.

\begin{table}[ht]
\caption{Dimensionless parameters for different electrode materials
  using the same geometry as in the experiment
  ($I=40$\,A, $R=44.5$\,mm, $h=11$\,mm, $\Delta T=8$\,K). For the
  material properties, see \cite{Sobolev2007,Sobolev2010,NEA2015,IAEA2008,Gale2004}.} 
\centering 
\begin{tabular}{c c c } 
\hline 
Component & A & B \\ [0.5ex] 
\hline 
  Na anode & 0.36 & 386\\
  PbBi cathode & 0.15 & 17.5\\
\hline
\end{tabular}
\label{table:1} 
\end{table}

\section{Results and discussion}\label{sec:results}
\begin{figure}[t!]
\centering
  \includegraphics[width=\textwidth]{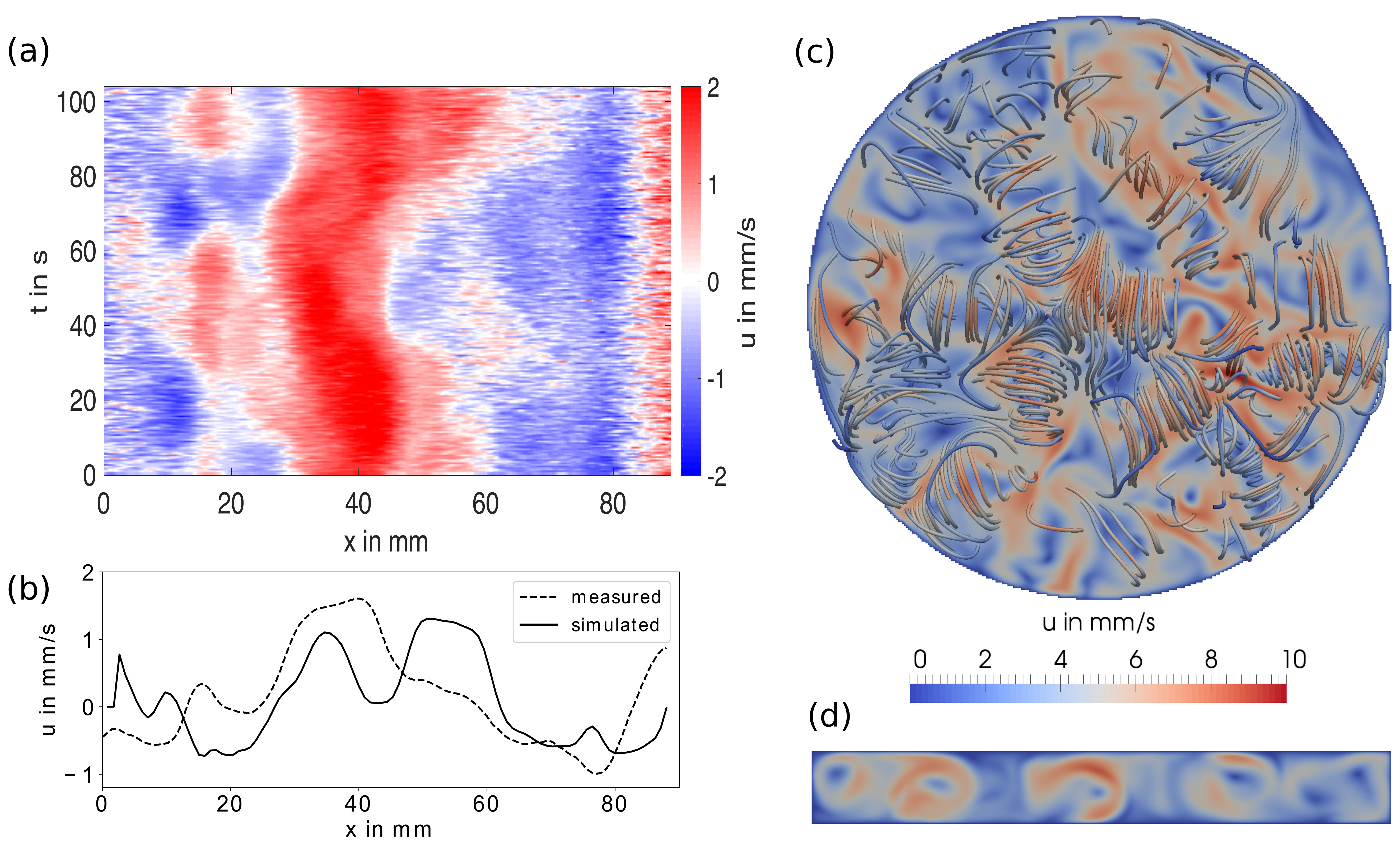}
\caption{Pure Rayleigh-B\'enard convection for $\Delta T=8$\,K. Measured radial velocity, (a). Mean speed across a diameter of the electrode, 6~mm above the vessel floor, (b), as measured (dashed curve) and simulated (solid curve). Speed and streamlines on a horizontal cross-section 5.5~mm above the vessel floor, (c). Speed on the vertical cross-section from which the ultrasound projection is extracted, (d). Rayleigh-B\'enard convection drives disordered flow.}
\label{fig:3}
\end{figure}

\begin{figure}[t!]
\centering
\includegraphics[width=\textwidth]{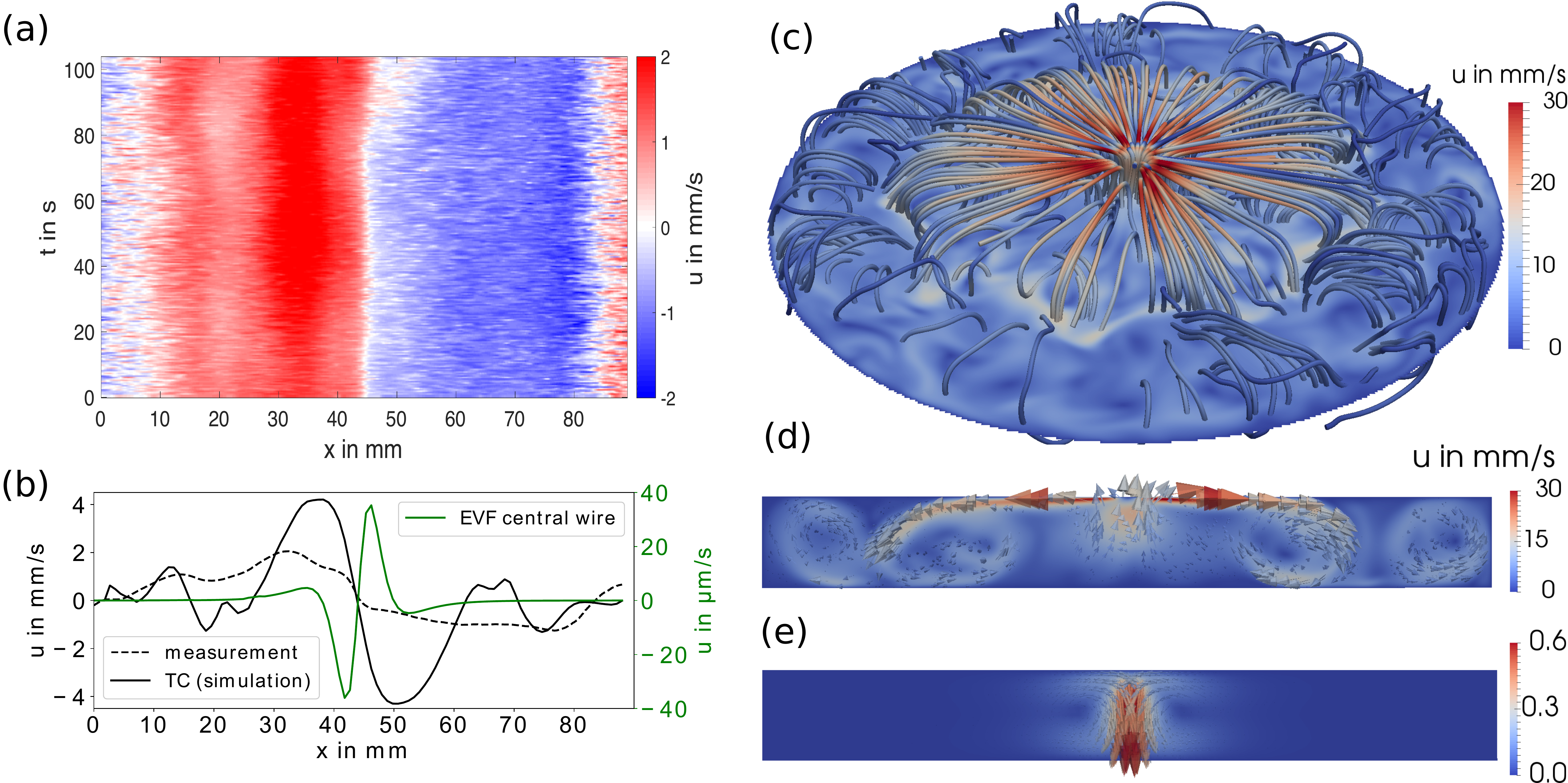}
\caption{Convection and electro-vortex flow with $I=2$~A. Measured radial velocity, (a). Mean speed across a diameter of the electrode, 6~mm above the vessel floor, (b), as measured (dashed curve) and simulated (solid curves). Speed and streamlines on a horizontal cross-section 5.5~mm above the vessel floor, (c). Flow due to internally heated convection (IHC) alone, on the vertical cross-section from which the ultrasound projection is extracted, (d). Flow due to electro-vortex flow (EVF) alone, on the same vertical cross-section, (e). IHC dominates EVF and has a shape consistent with experimental observations, but larger magnitude.}
\label{fig:4}
\end{figure}

\begin{figure}[t!]
\centering
  \includegraphics[width=\textwidth]{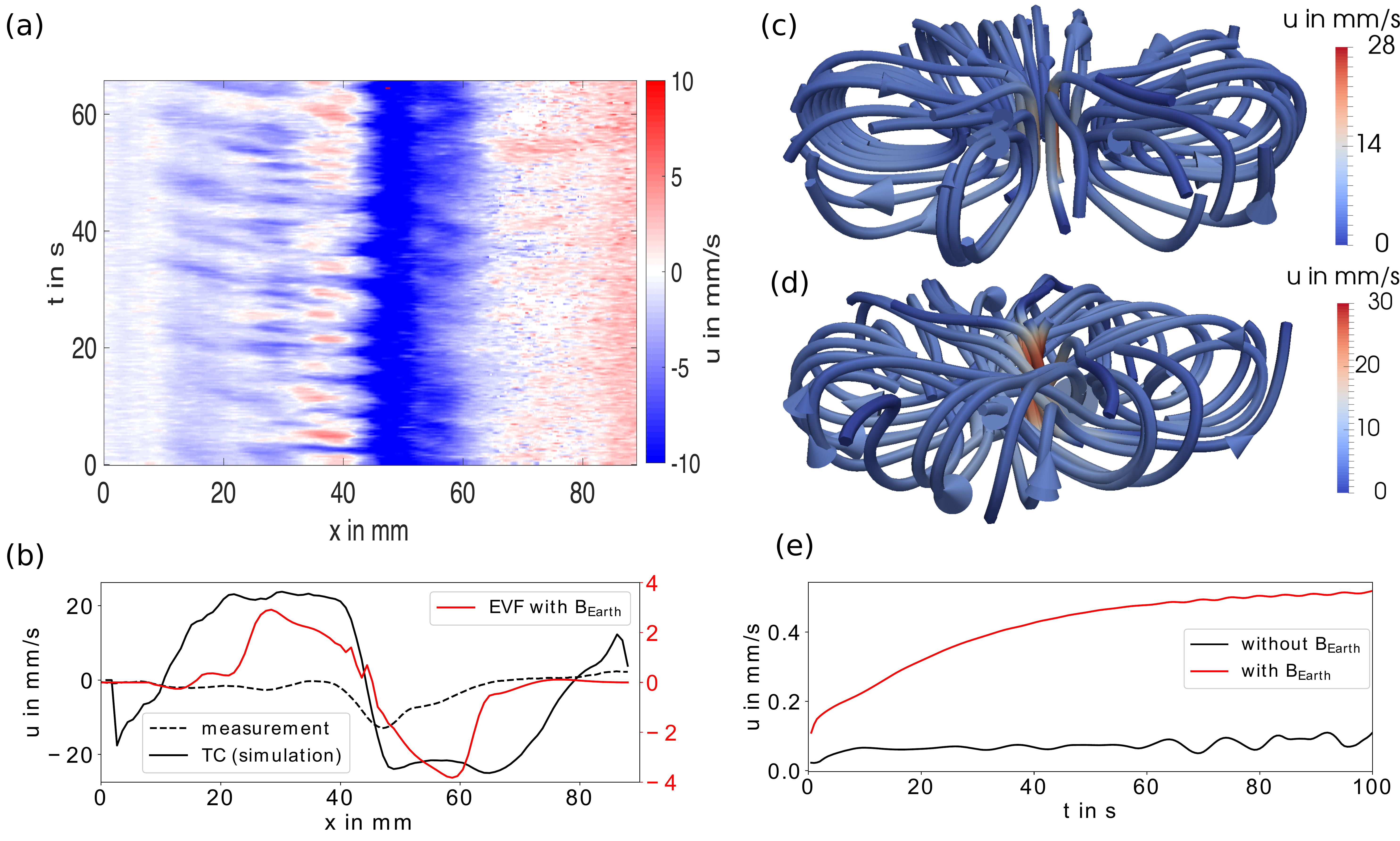}
  \caption{Convection and electro-vortex flow with $I=40$~A. Measured radial velocity, (a). Mean speed across a diameter of the electrode, 6~mm above the vessel floor, (b), as measured (dashed curve) and simulated (solid curves). Streamlines of electro-vortex flow simulated without Earth's magnetic field (c), and with it (d). Volumetrically-averaged poloidal and azimuthal velocity of electro-vortex flow (e). In (a) and (b), both internally heated convection and Rayleigh-B\'enard convection are present. Earth's field causes azimuthal flow in both experiments and simulations, but suppresses poloidal flow more strongly in experiments.}
\label{fig:5}
\end{figure}

In this section we present the measured and simulated velocities in
the liquid electrode without any current (Fig. \ref{fig:3}), with $I=2$~A
(Fig. \ref{fig:4}) and $I=40$~A (Fig. \ref{fig:5}). In all experiments,
velocity profiles are recorded when the temperature of the vessel is
stabilized at \SI{160}{\celsius}. The measured temperature difference
between the top and bottom of the liquid metal was always in the range
of $7$\,K$\le T-T_0 \le$~9\,K. 

Figure~\ref{fig:3}a shows the measured radial velocity distribution along the
diameter of the electrode recorded for two minutes. No current is
applied; we observe only Rayleigh-B\'enard convection. The flow is
rather disordered. This fits well to the numerical results
(Fig. \ref{fig:3}c), obtained with no-slip boundary conditions for
velocity, a fixed vertical temperature gradient and adiabatic side
walls. (The oxide film formed on the melt justifies using
a no-slip boundary condition for the free surface \cite{Cramer2014}.) Using the
ultrasound beam model described in \ref{s:beamModel} we
extract the radial velocity projection along the diameter of the UDV
probe so that we can directly compare the mean velocity from
experimental measurements with numerical simulations. Figure
\ref{fig:3}b shows the mean velocity from experimental and numerical
measurements (averaged over 100\,s). Both agree fairly well with respect
to the flow speed and the number of convection cells (illustrated in
Fig. \ref{fig:3}d). Though the flow is rather disordered, structure may be imposed via pinning of convection cells or heat transfer through the side walls. The top electrode certainly cools the
metal bath in the middle. Such a finger cooling is well known to produce an axisymmetric flow in the
Czochalski process of crystal growth~\cite{Cramer2014,Pal2015}. Our
configuration might also be compared to a flow driven by a localized
(external) heat source
\cite{Lopez2013,Navarro2007,Navarro2007a,Navarro2011,Navarro2013}.

When a (small) current of $I=2$~A is applied to the cell, the measured flow
becomes immediately much more ordered (compare Fig. \ref{fig:3}a and
\ref{fig:4}a); the measured flow magnitude does not change. Previous work dealing
with liquid metal convection in the presence of magnetic fields
suggest that convection rolls will tend to align with magnetic field
lines, producing a more ordered flow structure
\cite{Burr2002,Tasaka2016}. However, there are two other ways to
explain the increase in order much better: electro-vortex flow
and internally heated convection (IHC). Fig. \ref{fig:4}e shows the simulated
electro-vortex flow: a downward jet below the upper electrode,
exactly as expected. In the same volume, most of the Joule heat will
be generated. The strong radial temperature gradient forms lateral
jets and several convection cells (Fig. \ref{fig:4}d).
Comparing Figs.~\ref{fig:4}d and e, three things
are noteworthy: EVF and thermal convection have an opposing flow
direction, the magnitude of the thermal flow is one order of magnitude
larger than EVF, and both are fairly difficult to measure in our
experiments. In vacuum arc remelting the same flow 
directions are observed. However, electro-vortex flow reaches~---
compared to thermal convection~--- a ``similar size''
\cite{Davidson2000} of e.g. 30\% of the typical convection speed~\cite{Kazak2013}. A direct
comparison of the (time averaged) mean velocity projection along the
diameter of the aforementioned flows is shown in Fig.~\ref{fig:4}b. The simulated flow profile for IHC
 matches the measurement well (black curves). However, the
simulation shows velocities twice as large as the measured ones. We
believe this discrepancy may be explained partially by the simplified
boundary conditions (no-slip for velocity; fixed vertical temperature
difference; adiabatic side-walls). Moreover, an interaction between
the opposing EVF and thermal convection may change the flow structure,
too. Despite the described challenges of measurement and simulation
we can state: internally heated convection dominates the flow. 

When $I$ is increased to 40\,A, the measured velocities increase by a
factor of five (Fig. \ref{fig:5}a). We no longer observe a global
poloidal circulation, but a tornado-like flow structure emerges~--- a
phenomenon well known from former experiments
\citep{Woods1971,Bojarevics1983,Bojarevics1989}. Unfortunately, it is
not possible to measure the azimuthal flow speed with our current
experimental apparatus. Anyway, we know when swirl appears first: at 20\,A, if the current
is supplied symmetrically and at 10\,A if is supplied from the side. 
As mentioned in the introduction, the \emph{appearance} of azimuthal
forces can easily be explained: the internal radial currents interact
with a vertical stray magnetic field. This may be the Earth's magnetic field of
$B_z\approx 0.5$\,mT or with the words of Bojarevics: the ``slightest
deviation from symmetry'' leads to swirl \cite{Boyarevich1977}, because it produces
such fields \cite{Bojarevics1983,Bojarevics1989,Davidson2001}. The
often observed \emph{dominance} of the azimuthal flow is harder to explain. While
Millere assumes the kinetic energy of the poloidal flow to be
``partially transferred to the rotational motion'' \cite{Millere1980}, 
Shtern finds a supercritical bifurcation of the poloidal flow leading
to swirl \cite{Shtern1995}. Davidson et al. point out that the model of the latter
author is oversimplified. He explains the dominance of swirl as
``poloidal suppression'' \cite{Davidson1999}, which is in the sense of
\cite{Millere1980} and \cite{Bojarevics1983}. Specifically, he
describes that the swirling motion forces the poloidal flow into a
thin, dissipative Ekman layer where it is suppressed efficiently.

In Figs.~\ref{fig:5}c and d we show the electro-vortex flow, simulated
with and without the Earth's magnetic field. The latter we measured in
Dresden as $\bi B=(15\cdot\bi e_x, 5\cdot\bi e_y, 36 \cdot\bi
e_z)$\,$\mu$T. Figure~\ref{fig:5}e illustrates, that indeed the
azimuthal velocity increases due to the Earth's magnetic field. In
contrast to Davidson, we do not observe a strong poloidal suppression
(compare with Fig.~9 in \cite{Davidson1999}); investigation of that
discrepancy is underway. Finally, we compare measurement and
simulation in Fig.~\ref{fig:5}b. Obviously, the profile of the flow
matches well, while its magnitude does not. Simulating the effects of
thermal convection and EVF separately is important, but it is not
enough. Both phenomena interact: presumably the electromagnetically
driven swirl flow suppresses the thermal flow as described by Davidson et al.
\cite{Davidson1999}. A more detailed experimental and numerical study
of that interaction is planned for the near future.

A strong swirl flow may enhance mass transfer, but will also lead to
differences in the centripetal pressure along the interface. If strong
enough it will deform the electrode-electrolyte interface and may
eventually lead to a short circuit of the cell
\cite{Munger2008a,Horstmann2017}. By this means, swirling
electro-vortex flow could trigger interface instabilities in LMBs.

\begin{figure}[t!]
\centering
  \includegraphics[width=0.6\textwidth]{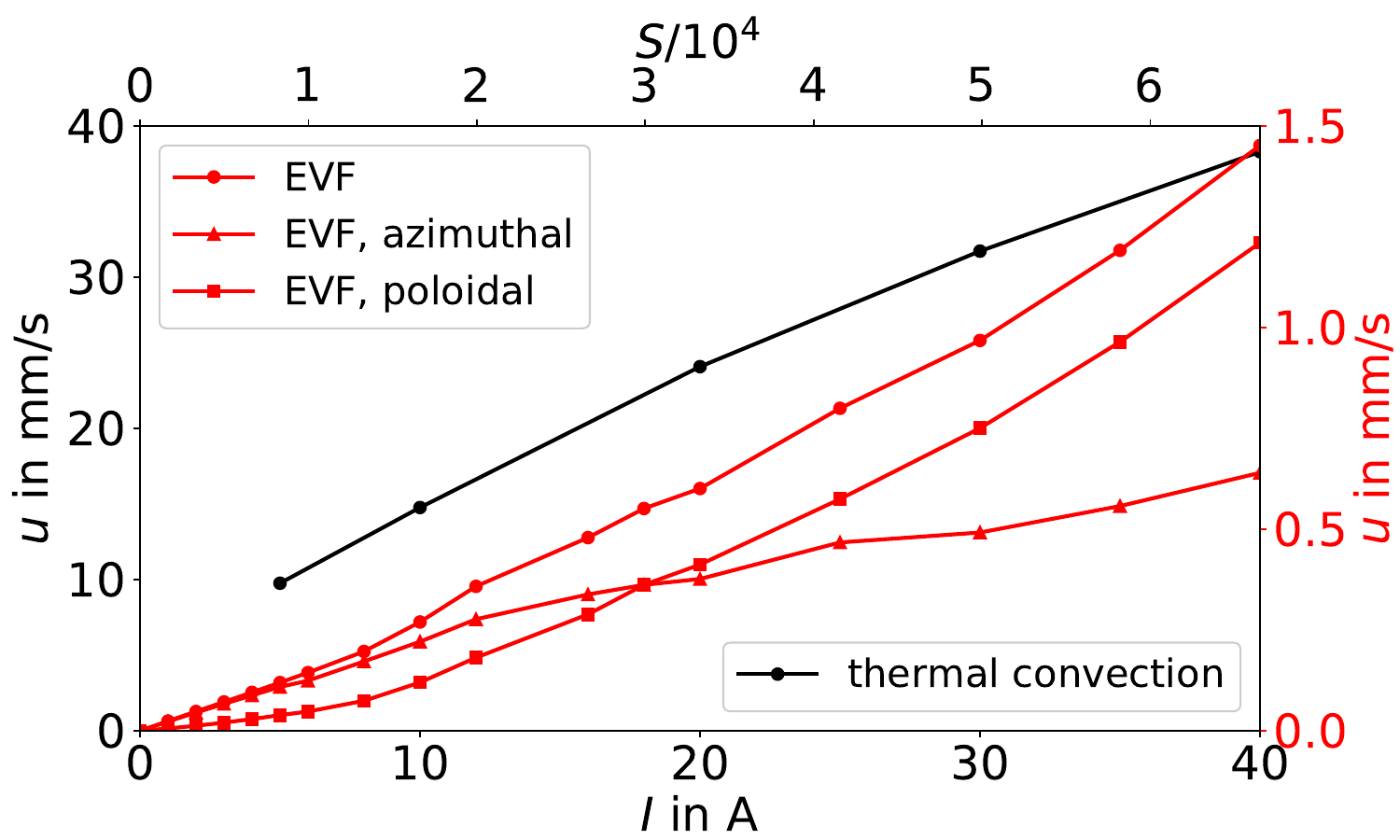}
  \caption{Volume-averaged velocity due to simulated internally heated convection (left axis) and electro-vortex flow (right axis). For these simulations we used symmetric current collectors and $\bi B=(15\cdot\bi e_x, 5\cdot\bi e_y, 36 \cdot\bi e_z)$\,$\mu$T. Though the speeds of both internal thermal convection and electro-vortex flow are predicted to scale linearly with current, we find that both deviate slightly.}
\label{fig:6}
\end{figure}

Figure \ref{fig:6} shows the simulated mean flow velocity of
internally heated convection (left axis) and electro-vortex flow
including the Earth's magnetic field (right
axis). As discussed before, thermal convection is one order of
magnitude larger than EVF. Though equation (\ref{eq:ScalingBuoyancy}) and (\ref{eq:dT}) predict that thermal convection speed will increase linearly with current, we find the rate of increase to be slower. On the other hand, the global EVF speed grows slightly faster than linearly
with current. According to Vlasyuk \cite{Vlasyuk1987b,Bojarevics1989}
we would expect a quadratic scaling up to $S=1\,000$ and above $S=10^5$ a linear scaling. Operating in the transition region between both, our results seem to fit roughly to Vlasyuks theoretical prediction.
Figure \ref{fig:6} also shows 
the poloidal and azimuthal mean velocity of EVF. The poloidal mean velocity
scales approximately quadratically with current, as one might expect. Increasing the cell
current, we also simultaneously increase the magnetic field induced by the current itself, with which the current interacts to drive poloidal flow. The rather linear
scaling of the azimuthal flow can be explained in a very similar
way. Here, only the cell current~--- together with the
\emph{constant} vertical field of the Earth~--- drives the flow.

\section{Summary, design implications, and outlook}
Using a combination of in-situ ultrasound velocity measurement and
numerical simulations, we studied the flow in a liquid PbBi
electrode. Specifically, we explored classical Rayleigh-B\'enard
convection, internally heated convection (IHC), electro-vortex 
flow (EVF) and swirl flow. Using a steel vessel and a point electrode
(Fig.~\ref{fig:2}) we applied up to 40\,A (0.65\,A~cm$^{-2}$). We 
demonstrated that meaningful UDV measurements are possible in a 
thin fluid layer of only 11\,mm height. For the
purpose of comparing experiment and simulation, we
developed a model for the ultrasound beam. Further, we derived
estimates of the relative importance of thermal convection and
electro-vortex flow, and pointed out the limitations of those estimates. 

We found the velocity of internally heated convection to be one order
of magnitude larger than electro-vortex flow. This relates to our
experiment (with the point electrode); it might be different in a real
liquid metal battery (LMB). In our experiment, IHC and EVF drive
flows of similar shape but in opposing directions. At low current
densities, we found thermal convection to dominate the flow
structure. The flow became more ordered with increasing current. At
higher currents we observed a strong interaction of all flow
phenomena. We found a good agreement of simulation and measurements for low
currents; as we modeled EVF and IHC separately (and did not account
for its interaction), experiment and simulation matched less closely at
higher currents. We found the mean velocity of EVF in typical LMBs
(10\,cm side length) to be in order of mm/s.

We further observed (azimuthal) swirl flow experimentally at 20\,A for
a symmetric experiment, or 10\,A for an asymmetric experiment. The numerical
simulations confirmed that vertical magnetic stray fields can cause
such flow. However, they did not show a suppression of the (original)
poloidal electro-vortex flow by the swirling motion. This is not in
line with the findings of \cite{Davidson1999}, probably due to the
different geometries. These differences deserve further
examination. We found the velocity of thermal convection to increase
less than linearly with the current. We explained why swirling motion
increases approximately linearly and poloidal electro-vortex flow
quadratically with current.

The locations and topology of electrical connections affect all four of the flow phenomena we have considered. The Joule heating that drives internally heated convection is a consequence of electrical current, so the placement and size of connections sets the IHC flow. Electro-vortex flow is driven by interactions between the current in the battery and its own magnetic field, both set by the connections with the electric circuit. Swirl flow is driven by interactions between current in the battery and magnetic fields produced by supply lines. And Joule heating of those supply lines can drive Rayleigh-B\'enard convection. 

Battery designers might choose locations and topology of electrical connections to promote mass transfer in the cathodes of liquid metal batteries, thereby allowing faster cycling and preventing formation of intermetallic phases. Rayleigh-B\'enard convection may be a suitable phenomenon to use for mixing. External temperature gradients may be imposed using the thermal management system of the cell stack (see \cite{Bradwell2015}). Internally heated convection can produce good mixing, but nearly all Joule heat in a liquid metal battery is produced in the electrolyte, located \emph{above} the cathode. The resulting stable temperature stratification tends to inhibit internally heated convection, not drive it. Thus IHC may also be problematic for designers. Electro-vortex flow seems more promising, because it can produce good mixing and because its magnitude can be adjusted by changing the diameter of the current collector to cause more or less divergence of current. Designers must understand, however, that they make a trade-off: reducing current collector size to promote mixing by electro-vortex flow also increases the resistance of the battery, reducing its voltage and efficiency. Also, real-world electro-vortex flows would likely be less vigorous than the ones in our experiments and simulations, which use a narrow current collector resulting in extreme divergence of current. We expect swirl to be less effective at promoting mass transfer than the other flow phenomena we have considered because it drives primarily horizontal motion, but vertical mass transfer is most helpful. We also expect swirl to deform the cathode-electrolyte interface substantially in large batteries. For that reason,
swirl flow will need special attention as it will surely trigger
interface instabilities. Designers should route battery supply wires so as to minimize vertical magnetic fields. 

Most importantly, the \emph{interaction} of internally heated
convection and electro-vortex flow deserves further study. In
addition, it will be important to reveal how the flow speeds of all
phenomena grow with larger currents in larger cells. The
appearance and growth of swirl deserves further numerical
investigation. In a real LMB cathode, EVF will drive a flow against a
stable temperature stratification. This configuration needs to be
studied, as well. Finally, flow speed measurements in a real 3-layer LMB
would be a great step forward.

\section*{Acknowledgments}
This work was supported by the National Science Foundation under award
number CBET-1552182, by the Deutsche Forschungsgemeinschaft (DFG,
German Research Foundation) under award number 338560565 as well as
the Helmholtz-Gemeinschaft Deutscher Forschungs\-zentren (HGF) in
frame of the Helmholtz Alliance  
``Liquid metal technologies'' (LIMTECH). The computations were
performed on the Bull HPC-Cluster ``Taurus'' at the Center for
Information Services and High Performance Computing (ZIH) at TU
Dresden and on the cluster ``Hydra'' at Helmholtz-Zentrum Dresden-Rossendorf. Fruitful discussions with A. Beltr\'an, V. Bojarevics,
P. Davidson, S. Franke, V. Galindo, G. Horstmann, J. Pal, J. Priede, D. R\"abiger,
F. Stefani, T. Vogt and T. Wondrak on several aspects of
electro-vortex flow and thermal convection are gratefully
acknowledged. N. Weber thanks Henrik Schulz for the HPC support. 

\section*{References}
\bibliography{literature}

\appendix
\section{Effect of acoustic beam broadening on UDV measurements}\label{s:beamModel}

In order to precisely compare experimental results gained from UDV
measurements to numerical simulations, it is necessary to account for
the fact that each ultrasound pulse broadens spatially as it travels
away from the emitting transducer, thereby reducing the lateral
resolution with increasing distance. Therefore, the measured velocity
at each distance interval will be a weighted average over the
radial fluid velocity component (pointing toward or away from the
transducer) within the width of the beam at that distance interval.

The pressure waves generated by a flat, circular transducer can be
thought of as the product of many adjacent point sources in accordance
with the Huygens-Fresnel principle. As a result, the ultrasound field
in the region near the transducer has a complicated structure. The
distance $z$ along the axis of the transducer and the intensity $I_z$
follow the relation 
\begin{eqnarray}
I_z =
I_0\sin^2\left(\frac{\pi}{\lambda}\left(\sqrt{r_T^2+z^2}-z\right)\right) 
\end{eqnarray}
rather than the inverse square law followed by the intensity of
spherical waves~\citep{Krautkraemer1986}. Here, $I_0$ is the maximum
intensity, $\lambda$ is the wavelength, and $r_T$ the radius of the
transducer. Along the transducer axis, intensity maxima occur when
\begin{eqnarray}
\frac{\pi}{\lambda}\left(\sqrt{r_T^2+z^2}-z\right) =
\left(\frac{1}{2}+n\right)\pi; n \in \mathbb{N}_0. 
\end{eqnarray}
The axial distances of these maxima are then
\begin{eqnarray}
z
=\frac{r_T^2-\left(\frac{1}{2}+n\right)^2\lambda^2}{2\left(\frac{1}{2}+n\right)\lambda}.
\end{eqnarray}
The furthest peak of this function occurs when $n =0$ at distance
\begin{eqnarray}
z_{nf} = \frac{r_T^2-\left(\frac{1}{4}\right)\lambda^2}{\lambda}.
\end{eqnarray}

This is the boundary of the near field, or Fresnel zone, beyond
which lies the far field, or Fraunhofer zone. In the Fresnel zone,
the acoustic field can be approximated as retaining the same cross
section as the transducer~\citep{Willemetz3000}. In the Fraunhofer
zone, the intensity decreases continuously (converging toward the
inverse square law at infinity). Moreover, the ultrasound field
spreads out conically. This can be quantified by the pressure wave's
directivity function $D(\theta)$, which weighs the acoustic
intensity with respect to the angle $\theta$ off the transducer axis
\begin{eqnarray}
D(\theta) = \frac{2J_1(kr_T\sin\theta)}{kr_T\sin\theta},
\end{eqnarray}
where $J_1$ is the Bessel function of the first kind and first order
and $k=\frac{2\pi}{\lambda}$ is the
wavenumber~\citep{Krautkraemer1986}. The intensity as a function of
the $z$ coordinate and angle is then 
\begin{eqnarray}
\frac{I(z,\theta)}{I_0} =
\sin^2\left(\frac{\pi}{\lambda}\left(\sqrt{r_T^2+z^2}-z\right)\right)(D(\theta))^2. 
\end{eqnarray}

The angle of divergence is where the intensity's first root lies~\citep{Krautkraemer1986,Willemetz3000}:
\begin{eqnarray}
\theta_0 = \arcsin\left(\frac{0.61\lambda}{r_T}\right).
\end{eqnarray}

$I(z,\theta)/I_0$ is used as a weighting factor when averaging
over the radial velocity components of each point within $\theta_0$ at
a certain distance from the transducer in the numerical model. This is done
to compare it to experimental results. The weighting factor and beam
broadening (green planes) are illustrated in Fig.~\ref{fig:2}.

\end{document}